# Data Based Linearization: Least-Squares Based Approximation

Zhentong Shao, *Student Member, IEEE*, Qiaozhu Zhai, *Member, IEEE*, Jiang Wu, *Member, IEEE*, and Xiaohong Guan, *Fellow, IEEE*

*Abstract*—Linearization of power flow is an important topic in power system analysis. The computational burden can be greatly reduced under the linear power flow while the model error is the main concern. Therefore, various linear power flow models have been proposed in literature and dedicated to seek the optimal approximation. Most linear power flow models are based on some kind of transformation/simplification/Taylor expansion of AC power flow equations and fail to be accurate under cold-start mode. It is surprising that data-based linearization methods have not yet been fully investigated. In this paper, the performance of a data-based least-squares approximation method is investigated. The resulted cold-start sensitive factors are named as least-squares distribution factors (LSDF). Compared with the traditional power transfer distribution factors (PTDF), it is found that the LSDF can work very well for systems with large load variation, and the average error of LSDF is only about 1% of the average error of PTDF. Comprehensive numerical testing is performed and the results show that LSDF has attractive performance in all studied cases and has great application potential in occasions requiring only cold-start linear power flow models.

## I. INTRODUCTION

POWER flow calculation is essential in power system operation and analysis. Although the conventional AC power flow (AC-PF) calculation yields accurate results, their non-linearity leads to computational obstacles in many optimization and control problems. Such as the difficulty in the convergence of optimal power flow computation [1] and the inconvenience in congestion analysis [2]. These disadvantages limit the application of AC-PF in system optimization, especially for large-scale systems [3]. Consequently, various linear power flow models [4-6] are proposed to relieve the computational burden caused by AC-PF. The linearization of power flow is beneficial for solving optimization problems because it allows the optimization problems to be transformed into linear programming problems.

Although the linearization of AC-PF provides computational advantages, it still brings approximation/model errors into power flow solutions [4]. Therefore, a well-developed linear power flow model would be attractive, provided it could offer sufficiently accurate power flow solutions. Besides, with the integration of renewable energy [7, 8] and the development of active distribution networks [9], power systems are more often in operating states with large perturbations (e.g., significant changes in bus power injections), and the traditional linear power flow models will likely break down more often in future power grids [10]. Therefore, it is important and necessary to develop a linearization method that has satisfactory approximation error and can work well in systems with large perturbations.

The essence of power flow linearization is to obtain the linear expression of branch power flow equations. In order to get a good linear expression, many linearization methods have been proposed. These methods can be roughly classified into two categories. They are 1) the DC power flow (DC-PF) model with its sensitivity form typically known as power transfer distribution factor (PTDF), and 2) the Taylor expansion method with its sensitivity form known as AC-PTDF.

The DC-PF model originates from engineering practice, and it has been widely used for a long time [11, 12]. The classical DC-PF model is derived from AC-PF equations by taking the assumptions of a lossless MW flow and a flat bus voltage profile. Detailed DC-PF options and classifications are summarized in [4], in which linear power flow models are divided into hot-start and cold-start models. Hot-start models correct the bus power injections according to a preset operating point from the AC-PF empirically [13]. This approach can improve the accuracy of linear power flow models locally. However, in applications such as system equivalent [3], transmission line planning [14], transmission constrained unit commitment (TCUC) [8] and local marginal price (LMP) fast calculation [15], no reliable AC-PF base point is available. Thus, the cold-start models like PTDF are indispensable in these applications. The PTDF based formulation is usually regarded as a large-signal sensitive power flow model, but the performance of PTDF is not robust enough as the maximum error of PTDF can reach hundreds of MW in large-scale systems [4]. This degree of numerical errors prevent it from being used in reliability related power system analysis.

The linearization methods based on Taylor expansion are prevalent in recent years [6]. These methods incorporate some assumptions adopted in DC-PF model and use the Taylor expansion of AC-PF equations at a specific operating point [16, 17]. The AC-PTDF is obtained by partial derivative (coefficients form the first-order Taylor expansion) of AC-PF equations [18]. These methods generally show excellent performance around the AC-PF base point, and they are more common in system real-time analysis [9]. In fact, the classic DC-PF model is mathematically the Taylor expansion of the AC-PF equation at an ideal operating point (lossless and nominal voltage). In this way, the two types of methods have some similar features. Due to their mathematical nature (Taylor expansion), these linearization methods tend to perform accurately near the base point, but not good enough when the operating point is far away from the base point.

Although the above methods do not guarantee a good enough linear approximation on a large variation of nodal injections, they all reveal the fact that there is a near linear relationship between the branch power flow and the nodal power injections. Therefore, a natural question is: what is the best linear approximation and how can we obtain this approximation?

Naturally, this idea reminds us about the mathematical concept of global optimal linear approximation. However, even for a function of only one variable, it is difficult to obtain the global optimal linear approximation. Then, are there any other good enough alternatives?

Before answering this question, the following problem is worthy of reflection.

*Should we establish a linear approximation that will be used **for all possible solutions of the AC-PF equations**, or a linear approximation only **for all the possible solutions that have either appeared in the past or will be appear in the future?***

For a given power system, let $AS$ be the set of all the possible solutions of the AC-PF equations, and let $RS$ be the set of all the possible solutions that have either appeared in the past or will be appear in the future. It is clear that $RS$ is a proper subset of $AS$. Therefore, the following features suggest that a linear approximation based on $RS$ is a good alternative to that based on $AS$.

- The approximation error of linear model on $RS$ is generally smaller than that on $AS$, especially when $RS$ is much smaller than $AS$;
- The set $RS$ can be obtained based on rich historical data. In this way, we don't even need to solve AC-PF equations to construct $RS$ when the system power flow states are included in the historical data;
- The computational complexity/burden of obtaining the linear power flow model will be greatly reduced when the problem is considered on $RS$, especially when only infinite number of power flow solutions are included in $RS$;
- The set $RS$ can be adjusted flexibly according to the engineering practice to give an adaptive optimal linear power flow model.

Altogether, it is intuitively reasonable to construct a linear approximation on the subset of power solutions that can appear in real operation rather than on the full solution set of the AC-PF equation.

Therefore, the motivation of this paper is to investigate the possibility and performance of a data-based linear power flow model. To this end, a new kind of cold-start sensitive factors named as least-squares distribution factors (LSDF) are obtained based on the historical data and thus a linearization method is established. Numerical testing is performed for several systems including a 2383-bus system and the results show that the performance of the proposed method is attractive. The approximation error of the LSDF based model is only about 1% of the traditional PTDF based model. Main features and generalization of the proposed method are also discussed.

The remainder of the paper is organized as follows: Section II provides the mathematical formulation of our method, which includes the symbology, methodology and discussion. Section III provides case study results and discusses the performance of the method. Section IV concludes the paper.

## II. MATHEMATICAL FORMULATION

### A. Global Optimal Linear Approximation Problem

This subsection proposes a general formulation for finding the global optimal linear approximation. The objective is minimizing the specified error metric between AC-PF model and the targeted linear power flow (LPF) model. The detailed formulations are as follows:

$$\min_{x} \max_{s} \ \text{error}(z^{AC\text{-}PF}(s), z^{LPF}(s;x)) \tag{1}$$
$$\text{s.t.} \qquad s \in AS(R)$$

Where,

$x$     denotes the linear factors that we seek.

$s$     denotes the power flow states/solutions/samples in set $AS$.

$z$     denotes the variable of interest (e.g., the branch power flow).

error(·)     denotes a specified error metric (e.g., the absolute or least-squares of approximation error).

$AS(R)$     denotes the set that includes all possible solutions of the AC-PF equations under a given system operating range. Generally, the system operating range can be defined as the system load variation range $R$. (e.g., $R$ = 40% means that the load vary from 60%~100% of the maximum load.

A straightforward way to solve problem (1) is decomposing the problem into main- and sub-problems as (2)-(3) and solve them iteratively. Actually, works in [10, 19, 20] have made good attempt in solving this iterative framework. We also reproduced these works and find the some difficulties:

**Main-Problem (MP):**

$$\min_{x} \ \text{error}(z^{AC\text{-}PF}(s), z^{LPF}(s;x)) \tag{2}$$
$$s.t. \quad s^{*} \in \text{worst scenario}$$

**Sub-Problem (SP):**

$$\max_{s} \ \text{error}(z^{AC\text{-}PF}(s), z^{LPF}(s;x)) \tag{3}$$
$$s.t. \qquad s \in AS(R)$$

*1) The computational burden is considerable.*

The iterative framework is computationally burdensome. Because **SP** is non-convex since it contains AC-PF equations. Non-convex programming problems need to be solved repeatedly during the iterative process. The convergence of **SP** cannot be guaranteed especially for large-scale systems, and the convergence of the iterative framework cannot be guaranteed either.

*2) Overestimate of approximation error in worst scenarios.*

$AS$ contains all possible solutions of AC-PF equations, but some of the solutions are meaningless in practice (e.g., power flow solutions with low voltage or high net losses). Considering these meaningless worst-scenarios in the framework will make the obtained LPF model perform poorly in other useful scenarios.

## B. Data-based Linearization Method

As the analysis in subsection A, seeking the global optimal linear approximation based on $AS$ is a behavior that 'pays a lot but returns a little'. In this subsection, we would like to discuss the possibility of obtaining a better LPF model based on $RS$.

$RS$ represents the most likely power flow states in real system operations. It can be defined by rich historical data of the system. The elements in $RS$ are called scenarios/samples. Each scenario contains all power flow information of the system at a specific time, including branch power flows, nodal power injections, and nodal voltage magnitudes and angles, etc. With $RS$, the problem (1) is transformed into (4):

$$\min_x \max_s \quad \text{error}(z^{AC\text{-}PF}(s), z^{LPF}(s;x)) \tag{4}$$
$$\text{s.t.} \qquad s \in RS$$

The problem (4) is very easy to solve because it is a pure linear programming problem. However, it is not a good alternative for problem (1) because it is also overestimate in worst-scenarios. The data quality of $RS$ directly affects the obtained LPF. An extreme scenario or a wrong data can lead to a completely wrong LPF.

$$\min_x \sum_{k=1}^{K} \left| \left( z^{AC\text{-}PF}(s^k), z^{LPF}(s^k;x) \right) \right|^2 \tag{5}$$
$$\text{s.t.} \qquad s^k \in RS, \ k = 1, 2, \cdots, K$$

### So how about the least-squares error metric in (5) ?

While retaining the advantage of easy-to-solve of problem(4), the least-squares method is not sensitive to the negative effects of worst-scenarios especially when the sample is sufficient. To illustrate the advantages of least-squares, a conceptual example of different linearization methods is shown in Fig. 1. In Fig. 1, the approximation effects of four methods are displayed intuitively. The slope of $DC\text{-}PF$ always reflects the trend of AC-PF model, but its intercept is empirical. The *Taylor-based* method is generally locally optimal. The *Min-Max* is affected by extreme points and inevitably overestimated. The *least-squares* guarantees a good linear approximation in most power flow solutions.

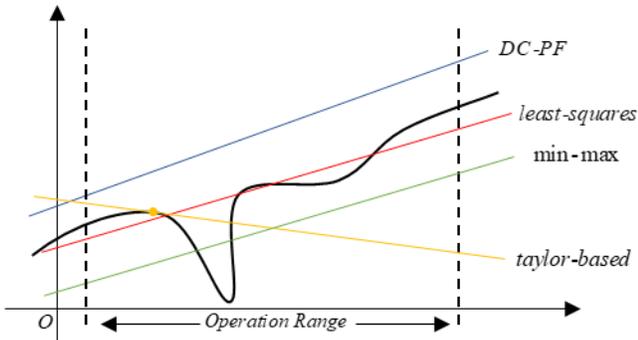

Fig. 1 Conceptual example of different linearization methods. Mathematically, the DC-PF is a trend line, the first-order Taylor approximation at a nominal operating point (golden dot) is a tangent line, and the least-squares and min-max methods are secant lines.

In the end, we chose the least-squares method as the research direction, because 1) it is pure linear programming and easy to solve, 2) it can guarantee a good linear approximation over a wide operation range, 3) it is adaptive in practice as $RS$ is defined based on actual systems' data, 4) it is adjusted flexibly by changing the data in $R$. We name this method as data-based linearization because the obtained LPF model is adaptive optimal to the data in $RS$.

## C. The Least-Squares Distribution Factors

Now we practice the above theory by a new kind of PTDF-liked LPF model. The model is named as least-squares distribution factors (LSDF). For sake of brevity, we define the following symbology:

$i$    is the index of system buses, $i = \{1, 2, \cdots, N\}$.

$l$    is the index of system branches, $l = \{1, 2, \cdots, 2L\}$. Note that one branch has two directions, so the total number of $l$ is $2L$.

$k$    is the index of samples in $RS$, $k = \{1, 2, \cdots, K\}$.

$P_i^{(k)}$    is the active power injection at bus $i$ under sample $k$.

$\mathbf{P}^{(k)}$    is the vector of active power injection under sample $k$, which is a $N \times 1$ vector.

$P_l^{(k)}$    is the branch active power at branch $l$ under sample $k$.

$\mathbf{P}_L^{(k)}$    is the vector of branch active power under sample $k$, which is a $2L \times 1$ vector.

$\mathbf{X}$    is the LSDF matrix, whose dimension is $2L \times N$.

$\mathbf{x}_l$    is the $l$-th row of the LSDF matrix, which is a $1 \times N$ vector.

$x_{l,i}$    is the $(l, i)$ element of the LSDF matrix.

We seek the optimal linear relationship between nodal active power injections $P_i$ and branch active power $P_l$. For a specific load variation range $R$, we gather samples from historical data and define the set $RS$. For each branch $l$, we form the optimization problem (6):

$$\min_{x_{l,i}} \sum_{k=1}^{K} \left| P_l^{(k)} - \sum_{i=1}^{N} x_{l,i} \, P_i^{(k)} \right|^2 \tag{6}$$
$$s.t. \qquad P_{l+}^{(k)}, P_i^{(k)} \in RS$$

We rewrite (6) into a matrix form:

$$\min_{\mathbf{x}_l} \quad f = \sum_k \left( P_{l+}^{(k)} - \mathbf{x}_l \mathbf{P}^{(k)} \right)^2 \tag{7}$$
$$s.t. \qquad P_{l+}^{(k)}, \mathbf{P}^{(k)} \in RS$$

With the derivation in (8), we get an unconstrained quadratic optimization problem in (9).

$$
\begin{aligned}
f &= \sum_k \left( P_l^{(k)} - \mathbf{x}_l \mathbf{P}^{(k)} \right)^2 \\
&= \sum_k \left\{ \mathbf{x}_l \left[ \mathbf{P}^{(k)} (\mathbf{P}^{(k)})^{\mathrm{T}} \right] \mathbf{x}_l^{\mathrm{T}} - 2 P_l^{(k)} (\mathbf{P}^{(k)})^{\mathrm{T}} \mathbf{x}_l + (P_l^{(k)})^2 \right\} \\
&= \mathbf{x}_l \left[ \sum_k \mathbf{P}^{(k)} (\mathbf{P}^{(k)})^{\mathrm{T}} \right] \mathbf{x}_l^{\mathrm{T}} - 2 \left[ \sum_k P_l^{(k)} (\mathbf{P}^{(k)})^{\mathrm{T}} \right] \mathbf{x}_l^{\mathrm{T}} + \sum_k (P_l^{(k)})^2.
\end{aligned} \tag{8}
$$

$$\min_{\mathbf{x}_l} f = \mathbf{x}_l \mathbf{A} \mathbf{x}_l^{\mathrm{T}} - 2 \mathbf{b}_l^{\mathrm{T}} \mathbf{x}_l^{\mathrm{T}} + c \tag{9}$$

Where,

$$\mathbf{A} = \sum_k \mathbf{P}^{(k)}(\mathbf{P}^{(k)})^T; \mathbf{b}_l = \sum_k P_l^{(k)}(\mathbf{P}^{(k)}); c = \sum_k (P_l^{(k)})^2 \quad (10)$$

The minimum value of $f$ is obtained when the gradient $\nabla f$ is equal to zero, which is:

$$\nabla f = 2\mathbf{A}\mathbf{x}_l^T - 2\mathbf{b}_l = 0 \Rightarrow \mathbf{A}\mathbf{x}_l^T = \mathbf{b}_l \quad (11)$$

The problem (11) can be solved in parallel for each branch $l$. It is worth attention that $\mathbf{A}$ is a $N \times N$ matrix and required to be non-singular. Even if $\mathbf{A}$ is not singular, the problem (6) can be solved by programming method. As the historical data is always rich, the matrix $\mathbf{A}$ is easy to be singular.

For each branch $l$, the matrices $\mathbf{A}$ are the same, so the problem (11) can be further integrated:

$$\mathbf{A}\mathbf{X}^T = \mathbf{B} \quad (12)$$

Where,

$$\mathbf{A} = \sum_k \mathbf{P}^{(k)}(\mathbf{P}^{(k)})^T$$
$$\mathbf{B} = (\mathbf{b}_1, \mathbf{b}_1, \cdots, \mathbf{b}_{2L}) \quad (13)$$

The solution of (12) is very fast. On a personal computer, it takes no more than 50 seconds even for a 2383-bus system. If the algorithm is executed in parallel, the solution time is no more than 10 seconds. In Fig. 2, we display the flow chart for obtaining LSDF in parallel.

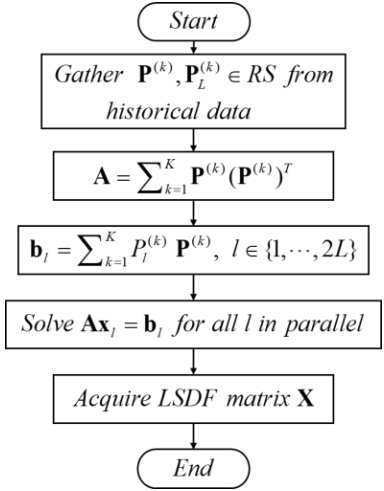

Fig. 2 The flow chart of the least-squares method

## D. Discussion and Outlook of Our Method

We now discuss the main features of LSDF and our data-based linearization method.

### 1) High accuracy of double-end LSDFs

It is worth attention that the data-based linearization method allows us to obtain the double-end distribution factors for all branches. The LSDF matrix $\mathbf{X}$ can be divided into $\mathbf{X}_+$ and $\mathbf{X}_-$. $\mathbf{X}_+$ represent the from-end distribution factors for all branches, and $\mathbf{X}_-$ represent the to-end distribution factors for all branches. In PTDF method, the from- and to-end distribution factors are same because the system is regarded as lossless. This double-end modeling method can effectively improve the accuracy of linear approximation. Let's illustrate this point with a simple example.

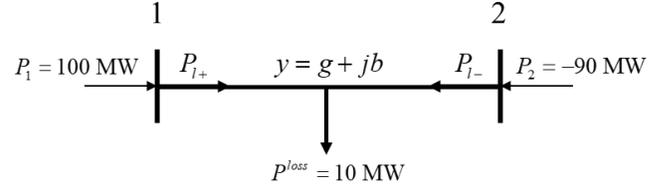

Fig. 3 2-bus system

There is a 2-bus and 1-branch system in Fig. 3. The power flow state is marked in the figure. Bus 1 is regarded as a slack bus when calculating PTDF.

According to the assumptions of DC-PF model, the PTDF of the 2-bus system can be easily obtained, which is $\Gamma = [0 \quad 1]$. The linear approximation results of PTDF are:

$$\left\{ \begin{array}{l} P_{l+} = 90 \text{ MW} \\ P_{l-} = -90 \text{ MW} \end{array} \right. \quad \left\{ \begin{array}{l} \text{Error}_{l+} = 10 \text{ MW} \\ \text{Error}_{l-} = 0 \text{ MW} \end{array} \right. \quad (14)$$

In contrast to PTDF, the approximation results of LSDFs are error-free, which are:

$$\left\{ \begin{array}{l} \mathbf{X}_+ = [1 \quad 0] \\ \mathbf{X}_- = [0 \quad 1] \end{array} \right. \quad \left\{ \begin{array}{l} P_{l+} = 100 \text{ MW} \\ P_{l-} = -90 \text{ MW} \end{array} \right. \quad (15)$$

The total loss can also be well modeled:

$$P^{loss} = P_{l+} + P_{l-} = 10 \text{ MW} \quad (16)$$

In fact, if the sum of nodal power injections is equal to the total system losses, LSDFs can give an estimate of losses at each branch and guarantee the total loss is error-free:

$$P_{\text{total loss}}^{\text{Approx}} = P_{\text{total loss}}^{\text{Real}} \quad (17)$$

This feature is proved in appendix.

### 2) There is no slack bus in LSDF

The inappropriate selection of slack bus will introduce the approximation errors to LPF models. The idea of distributed slack bus is proposed in [9] to relieve the bad effect of slack bus. On the contrary, there is no slack bus in the calculation of LSDF. This means that the information of all buses is considered, and this is one of the reasons for the high performances of our method. From another perspective, it can also be considered that LSDF contains slack buses, which are distributed slack buses that are adaptively assigned based on data in *RS*.

### 3) Discussion for obtaining samples of RS

We recommend historical data as samples in *RS*. Because 1) historical data can well reflect the true operation states of practical power systems, and 2) the power grid corporation has accumulated a large amount of power flow data but cannot use the data effectively. The obtained data needs to be in the same system topology. Once the topology changes, the LSDF needs to be recalculated, which is the same as PTDF. In addition, when data is sufficient, the least-squares method will not be affected by a small amount of wrong data, and the requirements for data quality are not harsh. Theoretically, the larger the amount of data, the better the LSDF performance. But in practice, it is found that the LSDF obtained by an appropriate amount of data can be very similar to the LSDF obtained on a huge amount of data. In the specified data set *RS*, LSDF will always perform better than PTDF, because PTDF is also one of the feasible solutions to problem (6).

### 4) The extension of data-based linearization

The data-based linearization is a general idea and LSDF is one of the concrete way of implementing this idea. This idea can be applied to obtaining various types of LPF models. Moreover, this method has great application potential in many power optimization problems, the attractive performance of the method can help improve optimal power flow (OPF), congestion analysis or other problems that require an accurate linear approximation.

### 5) Error bound on LSDF in future work

The linear power flow models LSDF performs well over a large number of Monte Carlo tests, but we still want to figure out the maximum error bound of the LSDF. In future work, we are dedicated to developing a general method for estimating the error bound of LPF models and making our approach be more complete.

## III. CASE STUDY

### A. Basic Information

Several IEEE test systems form MATPOWER 7.0 [21] are tested. Simulation samples are used instead of historical samples from actual systems. MATPOWER provides the maximum load value of each system, so we generate samples by multiplying the maximum load with a random coefficient. Here are the detailed steps:

1) Given the max-load of each load bus.

2) Create a random coefficient uniformly within a specific range.

3) Modifying the load with (18).

4) Solving power flow equations under the load.

$$\begin{cases} P_{load,i}^{(k)} = P_{load,i}^{\max} \times \eta_A \times \eta_i^P; i = 1, 2, \cdots, N \\ Q_{load,i}^{(k)} = Q_{load,i}^{\max} \times \eta_A \times \eta_i^Q; i = 1, 2, \cdots, N \\ \eta_A \in [1 - R, 1]; \ \eta_i^P, \eta_i^Q \in [0.95, 1.05]. \end{cases} \quad (18)$$

The random coefficients $\eta_A$, $\eta_i^P$ and $\eta_i^Q$ are generated uniformly. Coefficients $\eta_A$ is used to adjust the overall load level. Coefficients $\eta_i^P$ and $\eta_i^Q$ is aimed at keeping $P_i$ and $Q_i$ from changing consistently. $R$ defines the load variation range of the system. The number of samples in $RS$ is another important indicator, so we use $RS(R,K)$ to define the sample set. For the convenience of expression, the total sample number $K$ is expressed as a multiple of the total number of system buses (e.g., $K = 10N$). The sample set for testing and the sample set for calculating LSDF are generated independently.

The comparison is carried out between the traditional PTDF method and the LSDF method. They are all used for used for systems with large perturbations or cold-start occasions.

### B. The Comparison of PTDF and LSDF

For practical power systems, the annual load variation range is generally less than 50% of max-load. So we provide three cases for comparison, which are: 1) CASE_I: $R = 20\%$, $K = 10N$; 2) CASE_II: $R = 40\%$, $K = 20N$; and 3) CASE_III: $R = 60\%$, $K = 30N$. Small range set are included in large range set (i.e., $RS^I \subset RS^{II} \subset RS^{III}$). It should be noted that for 300-bus, 1354-bus, the maximum $R$ that system can withstand is only about 30%. At this time, we revise the cases as CASE_I: $R = 10\%$, $K = 10N$; 2) CASE_II: $R = 20\%$, $K = 20N$; and 3) CASE_III: $R = 30\%$, $K = 30N$.

Table I shows the overall performance of PTDF and LSDF in several systems. 24-bus, 118-bus, and 2383-wp systems are selected as representatives for displaying the detailed error information (The 'Form-direction' and 'To-direction' branch power flows are unified for display), which are shown in Fig. 4, Fig. 5 and Fig. 6. Form these test results, we find the follows:

1) The approximation error of LSDF is much smaller than that of PTDF. In Table I, it is found that the total Avg. Err of LSDF is only 0.71% of that of PTDF, and the total Max. Err of LSDF is only 1.26% of that of PTDF. This means the approximation error of LSDF is only about 1% of PTDF.

TABLE I
COMPARISON RESULTS OF LSDF AND PTDF ON SEVERAL SYSTEMS

| Test System | CASE_I $R = 20\%$ | | | | CASE_II $R = 40\%$ | | | | CASE_III $R = 60\%$ | | | |
| --- | --- | --- | --- | --- | --- | --- | --- | --- | --- | --- | --- | --- |
| | Avg. Err (MW) | | Max. Err (MW) | | Avg. Err (MW) | | Max. Err (MW) | | Avg. Err (MW) | | Max. Err (MW) | |
| | LSDF | PTDF | LSDF | PTDF | LSDF | PTDF | LSDF | PTDF | LSDF | PTDF | LSDF | PTDF |
| 5-bus | 0.014 | 0.856 | 0.073 | 2.184 | 0.015 | 0.886 | 0.074 | 2.528 | 0.015 | 0.892 | 0.074 | 2.930 |
| 24-bus | 0.044 | 4.871 | 0.503 | 43.515 | 0.055 | 5.177 | 0.606 | 46.655 | 0.063 | 5.542 | 0.687 | 46.655 |
| 30-bus | 0.009 | 0.578 | 0.105 | 10.264 | 0.009 | 0.631 | 0.105 | 10.264 | 0.010 | 0.689 | 0.130 | 10.264 |
| 57-bus | 0.016 | 1.360 | 0.160 | 6.693 | 0.018 | 1.385 | 0.260 | 9.187 | 0.022 | 1.402 | 0.285 | 9.187 |
| 118-bus | 0.018 | 3.408 | 0.891 | 45.455 | 0.021 | 3.502 | 1.200 | 45.455 | 0.027 | 3.662 | 2.591 | 58.863 |
| 300-bus | 0.084 | 10.552 | 6.103 | 435.863 | 0.088 | 10.928 | 8.579 | 441.422 | 0.103 | 11.625 | 8.579 | 451.271 |
| 1354-bus | 0.033 | 8.735 | 1.327 | 341.443 | 0.034 | 8.735 | 1.327 | 358.388 | 0.036 | 8.802 | 2.173 | 425.802 |
| 1888-bus | 0.017 | 4.387 | 1.141 | 136.443 | 0.019 | 4.401 | 1.917 | 137.479 | 0.026 | 4.585 | 3.594 | 197.491 |
| 2383-wp | 0.005 | 2.345 | 0.334 | 109.258 | 0.007 | 2.345 | 1.086 | 146.170 | 0.011 | 2.354 | 2.383 | 146.170 |
| Total | 0.240 | 37.092 | 10.637 | 1131.12 | 0.266 | 37.99 | 15.154 | 1197.55 | 0.313 | 39.553 | 20.496 | 1348.63 |

Analyzing test results of each system in Table I, it can be found that Max. Err of LSDF is almost no more than 10 MW, but the Max. Err of PTDF is generally tens of MW, and in large test systems, it even reaches hundreds of MW, e.g., the Max. Err of PTDF in 300-bus system is 451.271 MW.

*2) Detailed analysis of the branch where the maximum error appears.* It is found that the errors trend to be larger in large-scale test systems. However, the approximation error does not have a proportional relationship with system's scale. This shows that some systems may have a high inherent linearity, while others have a low linearity due to their own network characteristics. The maximum error of LSDF and PTDF both appears at the 300-bus test system under CASE_III. For more details, the Max. Err of LSDF appears at branch 177, which is a transformer branch connecting to a generator bus. The actual power value of branch 177 is 618.17 MW, and the approximation of LSDF is 609.59 MW. The error value is 8.579 MW, and the error percentage is 1.39%. In addition, it is found that in most test systems (except for 5-bus system because there is no transformer branch marked), the maximum error of LSDF almost appears at transformer branches. This phenomenon may indicate that the approximation error is related to reactive power, but LSDF has not considered the reactive power yet. On the other hand, the Max. Err of PTDF in 300-bus system appears at branch 403, which is a transformer branch connects to the slack bus. The actual power of branch 403 is 1407.87MW. The approximation of PTDF is 956.60 MW. The error value is 451.27 MW, and the error percentage is 32.05%. In other test systems, the maximum error of PTDF also appears near the slack bus. This phenomenon confirms the concern in literature [4, 9]. The selection of slack bus does bring model errors to PTDF.

*3) Influence of load variation range on performance of LSDF and PTDF.* It is found that as the load variation range

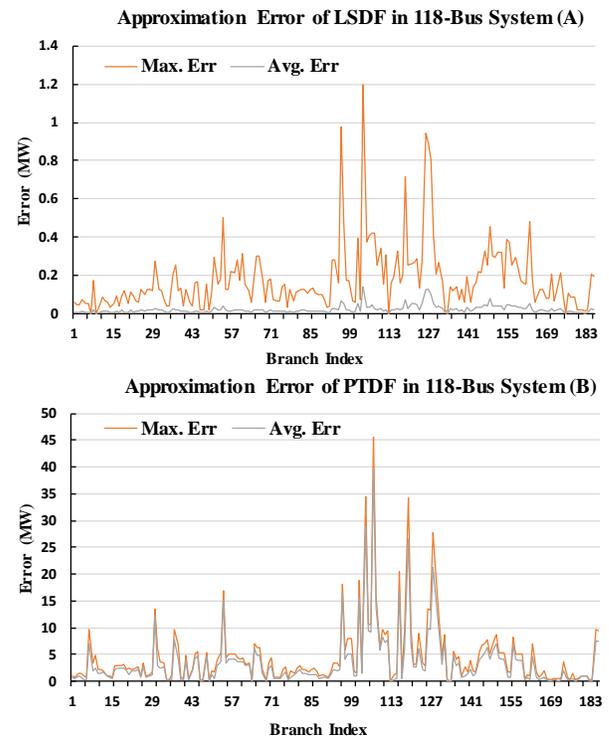

Fig. 5 Max. Err and Avg. Err of each branch in 118-bus system

expands, the approximation error of both LSDF and PTDF increases. From *R*=30% to *R*=70%, the average approximation error of LSDF increases from 0.240 MW to 0.313 MW. Although the error percentage increases by 30.4%, the error value only increases a little. This phenomenon shows that the performance of LSDF is relatively stable, and it can indeed be applied in systems with large perturbations. As for PTDF method, from *R*=30% to *R*=70%, the average approximation error increases from 37.092MW to 39.553 MW, and the error percentage increases by 6.63%. The performance of PTDF is basically in our expectations. Its approximation performance is also very stable, and it can also be applied in system with large perturbations.

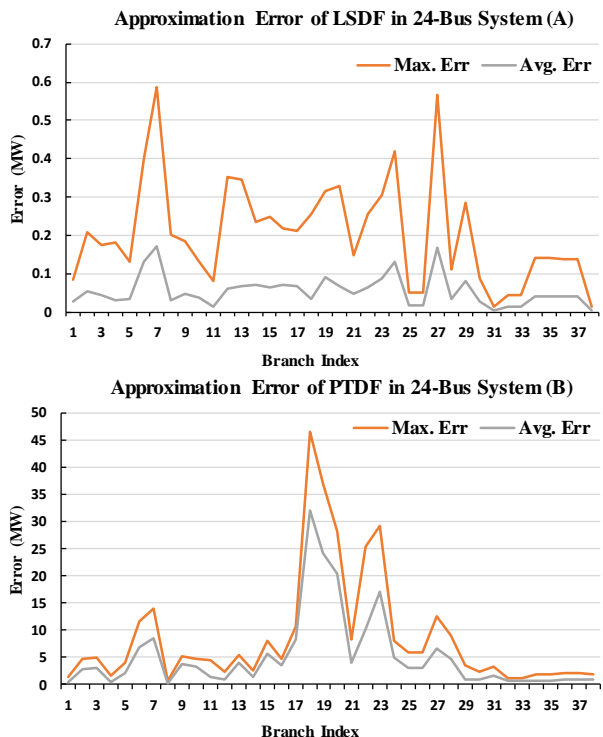

Fig. 4 Max. Err and Avg. Err of each branch in 24-bus system

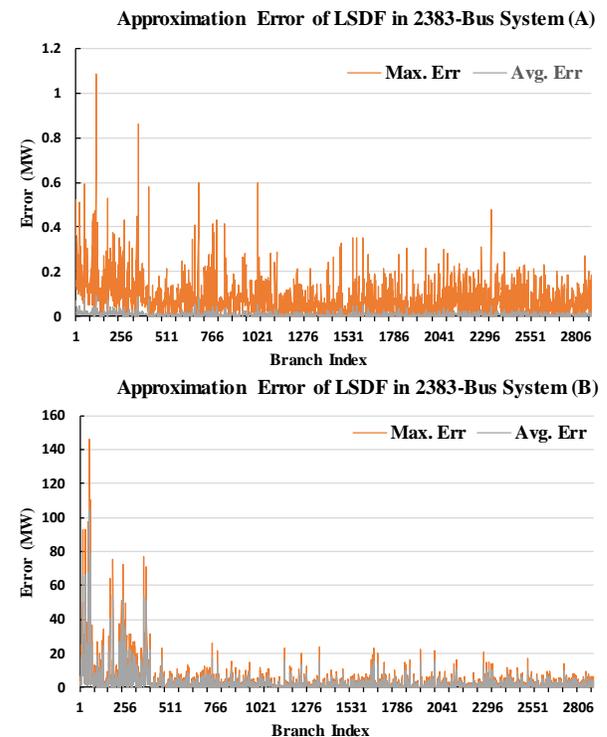

Fig. 6 Max. Err and Avg. Err of each branch in 2383-bus system

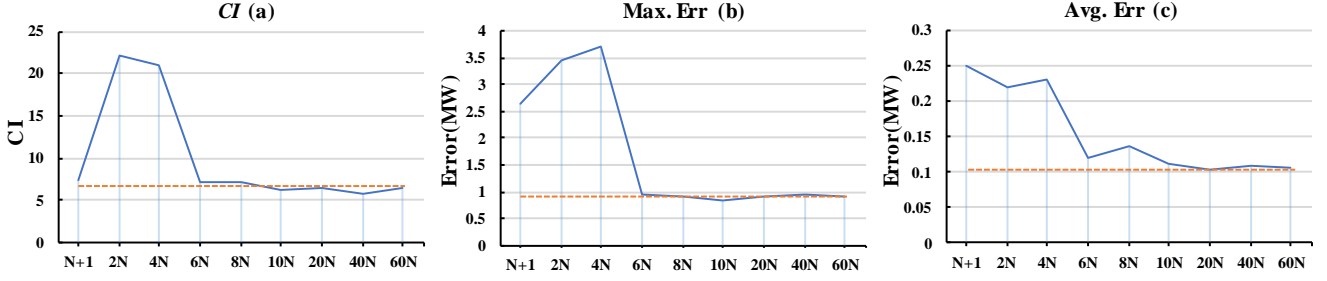

Fig. 7 The convergence results on 5-bus system (the abscissa of (a)-(c) is the number of selected samples, the orange dot line is the *CI* of *AS*)

4) *Detailed information of the error at each branch*. Detailed error information in CASE_II is shown in Fig. 4, Fig. 5 and Fig. 6. It is found that the Max. Err and Avg. Err are relatively far apart in LSDF. That is to say the LSDF guarantees high performance in most test samples, and only performs poorly in a few worst-samples. This phenomenon confirms the conceptual example of least-squares in Fig. 1. It shows that the LSDF effectively avoids the overestimation phenomenon and ensures the approximation accuracy in most samples over a large range. Besides, it is worth mentioning that the approximation error of LSDF in the worst samples is still very small in value. On the contrary, the approximation performance of PTDF in Max. Err and Avg. Err is relatively consistent and close. That is to say the approximation performance of PTDF in local worst samples is the same as that in most general samples. The reason for this phenomenon can also be found in Fig.1. It is shown in Fig. 1 that when the DC-PF is far from the AC-PF, the difference between worst points and general points can be ignored. In Fig. 4(B)-6(B), it is found that there always be a maximum error peak in approximation error of PTDF (e.g., $l = 18$ in Fig. 4(B), $l = 107$ in Fig. 5(B), and $l = 58$ in Fig. 6(B)). Checking for detailed branch parameters, it is found that these branches are all transformer branch directly connected to the slack bus. In Fig. 4(A)-6(A), it is found that there are several similar error peaks in LSDF. After checking the branch parameters, it is found that most of these peaks appear at transformer branches. The reason for this phenomenon are: 1) the transformer branches connected to generators often have a large power flow value, it is natural for a branch with large power flow value to have a large error value; 2) the transformer branches are different form the transmission branches in parameters (standard $\pi$ model). For example in 118-bus, the average transmission parameters are (19) and the average transformer parameter are (20). From the parameters, it can be seen that the reactive power flow through the transformer branch is relatively large, so it is speculated that the approximation error of the branch is related to the fact that the reactive power information is ignored.

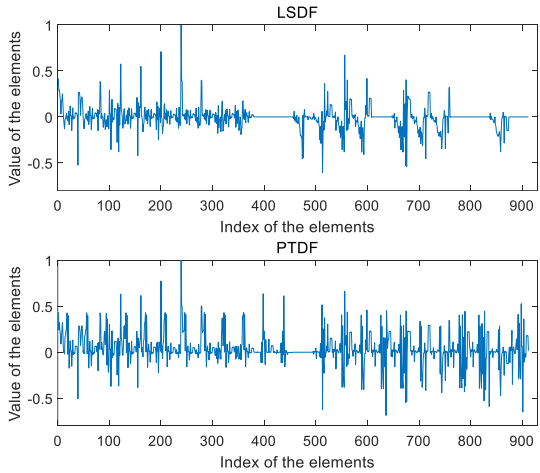

Fig. 8 The value of elements of LSDF and PTDF.

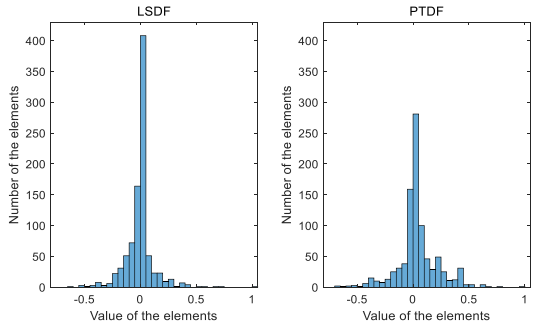

Fig. 9 The distribution of elements of LSDF and PTDF.

$$r_l + jx_l = 0.029 + j0.110, b_l = 0.075; \quad x_l / r_l = 3.8 \quad (19)$$

$$r_l + jx_l = 0.0026 + j0.0481, b_l = 0.019; \quad x_l / r_l = 18.5 \quad (20)$$

Where, $r_l$ is the resistance of branch $l$, $x_l$ is the reactance of branch $l$, and $b_l$ is the ground susceptance of branch $l$.

### C. The Convergence of LSDF

A very natural question is, how many samples are required for a good LSDF? We performed a test with 125000 samples in 5-bus system. There are 3 load buses in 5-bus system, and at each load bus, we obtain 50 load values evenly within the specified range $R = 50\%$ at even intervals. Enumerating all combinations, and finally obtain 125000 ($50^3$) scenarios. We assume this set as the *AS*. Then we select a certain number of samples in *AS* and calculate the corresponding LSDF matrix. Comparing these obtained LSDF matrices with the LSDF matrix obtained using all samples in *AS*. We finally get the results shown in Fig 7. The indicator *CI* is used to indicate the value of *AS*. It is defined as:

$$CI = \frac{1}{2} \sqrt{\sum_{i=1}^{N} \sum_{l=1}^{2L} (x_{l,i})^2} \quad (21)$$

In Fig 7, it is found that the LSDF converges when the number of samples reaches $10N$. That is to say, just selecting 50 samples is enough to obtain a LSDF matrix that is close to the global optimal LSDF matrix and performs quite well.

### D. The Physical Meaning of LSDF

The PTDF matrix characterizes the ratio of allocating nodal power injections to branches. We compare the elements of LSDF and PTDF in 24-bus system, and show the results in Fig.8 and Fig.9. In Fig.8, the elements of LSDF and PTDF are displayed in the same order. From the figures, it is found that the maximum coefficients of LSDF and PTDF are both 1, and the coefficient distributions of them are generally similar. The above results indicates that LSDF not only performs well, but also has the same physical meaning as PTDF.

## IV. CONCLUSION

This paper presents a data-based linearization method and makes a meaningful investigation on this method. Based on this idea, a new type of PTDF-liked linear power flow model is proposed in this paper. The LSDF, which is the proposed linear power flow model, has excellent approximation performance and has physical meanings close to PTDF. The proposed method has great potentials in many power optimization problems, especially when the systems are operating under large perturbations or systems require cold-start linear models only. This method has data-based adaptability and it is easy to be extended and applied to other types of linear power flow models. Hope this article can provide new ideas for the field of power flow linearization.

## APPENDIX

The approximation of the active power flow on lines are as follows:

$$P_{l+}^A = \mathbf{x}_{l+}\mathbf{P} \tag{22}$$

$$P_{l-}^A = \mathbf{x}_{l-}\mathbf{P} \tag{23}$$

Where, Superscript 'A' denote the approximation result, and superscript 'R' denote the real variable value.

**Proof:** First, we prove (24) ($\mathbf{e}$ is a $N \times 1$ vector whose elements are all equal to 1).

$$\sum_{l=1}^{L}\left(\mathbf{x}_{l-}^T + \mathbf{x}_{l+}^T\right) = \mathbf{e} \tag{24}$$

The proof of (24) is:

$$
\begin{aligned}
\mathbf{A} \cdot \mathbf{e} &= \left\{\sum_{k=1}^{K}\mathbf{P}^{(k)}(\mathbf{P}^{(k)})^T\right\} \cdot \mathbf{e} \\
&= \sum_{k=1}^{K}\mathbf{P}^{(k)}[(\mathbf{P}^{(k)})^T\mathbf{e}] = \sum_{k=1}^{K}\mathbf{P}^{(k)}[\sum_{i=1}^{N}P_i^{(k)}] \\
&= \sum_{k=1}^{K}\mathbf{P}^{(k)}[\sum_{l=1}^{L}P_{l+}^{(k)} + P_{l-}^{(k)}] \\
&= \sum_{l=1}^{L}\left(\sum_{k=1}^{K}P_{l+}^{(k)}\mathbf{P}^{(k)} + \sum_{k=1}^{K}P_{l-}^{(k)}\mathbf{P}^{(k)}\right) \\
&= \sum_{l=1}^{L}\left(\mathbf{b}_{l+} + \mathbf{b}_{l-}\right) = \sum_{l=1}^{L}\left(\mathbf{A}\mathbf{x}_{l-}^T + \mathbf{A}\mathbf{x}_{l+}^T\right) \\
&= \mathbf{A} \cdot \sum_{l=1}^{L}\left(\mathbf{x}_{l+}^T + \mathbf{x}_{l+}^T\right)
\end{aligned} \tag{25}
$$

So $\mathbf{e} = \sum_{l=1}^{L}\left(\mathbf{x}_{l+}^T + \mathbf{x}_{l+}^T\right)$ and (24) is proved.

Second, combining (22)-(23) and (24), we have:

$$
\begin{aligned}
P_{\text{total loss}}^A &= \sum_{l=1}^{L}\left(P_{l+}^A + P_{l-}^A\right) \\
&= \sum_{l=1}^{L}\left(\mathbf{x}_{l+}\mathbf{P} + \mathbf{x}_{l-}\mathbf{P}\right) \\
&= \left[\sum_{l=1}^{L}\left(\mathbf{x}_{l+} + \mathbf{x}_{l-}\right)\right]\mathbf{P} \\
&= \mathbf{e}^T\mathbf{P} = \sum_{i=1}^{N}P_i = P_{\text{total loss}}^R
\end{aligned} \tag{26}
$$

Q.E.D